\documentclass[aps,twocolumn,prl]{revtex4}

\begin{document}

\title{Comment on ``Ambiguities in the Up-Quark Mass''}

\author{Mark Srednicki}
\email{mark@physics.ucsb.edu}
\affiliation{Department of Physics, University of California,
Santa Barbara, California 93106, USA}


\maketitle

In a recent Letter, Creutz \cite{cr} argued that instanton effects in quantum chromodynamics lead to an additive ambiguity in the definition of the light quark masses, and that this ``calls into question the acceptability of attempts to solve the strong $CP$ problem via a vanishing mass for the [up] quark.''   Here we show that, contrary to this claim, the instanton effects discussed in \cite{cr}
actually enhance (rather than interfere with) the viability of the $m_u=0$ solution of the strong 
$CP$ problem.

To better understand the role of these instanton effects, we must treat the $CP$ violating phase 
$\theta$ explicitly; this was not done in \cite{cr}.  We begin by setting the coefficient of the
topological term $F\tilde F$ in the lagrangian to zero; we then identify $\theta$ as the phase 
of the determinant of the light-quark mass matrix $m$.  Instanton effects on the renormalization 
of $m$ can be accounted for explicitly via an extra term \cite{gm,choi,bns} in the 
renormalization-group equation for $m$,
\begin{equation}
a{d\over da}m = \gamma(g)m + c(g)a^{n_f-2}(\det m^\dagger)(m^\dagger)^{-1},
\label{dmda}
\end{equation}
where $a$ is the short-distance cutoff, $n_f$ is the number of light flavors, and
$c(g)=c_0(8\pi^2\!/g^2)^6 e^{-8\pi^2\!/g^2}[1+O(g^2)]$; $c_0=0.048$ for $n_f=3$.
Although the coefficient $c(g)$ in Eq.~(\ref{dmda})
is found via an instanton calculation, the $m$ dependence of this term
is fixed by the transformation properties of $m$ under the chiral flavor group
$SU(n_f)\times SU(n_f)$.  

If we left-multiply Eq.~(\ref{dmda}) by $m^{-1}$, take the trace, and use
$d \det m=(\det m)\,{\rm Tr}\,m^{-1}dm$, we find
\begin{eqnarray}
a{d\over da}\det m &=& n_f\gamma(g)\det m 
\nonumber \\
&& {} + c(g)a^{n_f-2}(\det m^\dagger m)\mathop{\rm Tr}(m^\dagger m)^{-1}.\quad
\label{ddetm}
\end{eqnarray}
From Eq.~(\ref{ddetm}), 
we see that the nonperturbative contribution to the renormalization
of $\det m$ is always real.  Thus, if $\det m$ vanishes at any particular scale $a$, 
it is real at all scales.  If we can explain why $\det m=0$ at any one scale
then we will have solved the strong CP problem.   

Models with $\det m=0$
at a high scale have been proposed, and involve spontaneous breaking of a ``horizontal'' or
``family'' symmetry; a general analysis of this class of models was given in \cite{bns}.
Nonperturbative contributions to ${\rm Re}\det m$ 
actually {\it improve\/} the status of these models \cite{gm,choi,bns}, 
because a nonzero up quark mass at
$1/a\sim \Lambda_{\rm\scriptscriptstyle QCD}\sim 1\;\rm GeV$ 
is generated by instanton effects at shorter distances.   This is 
different than the superficially similar Kaplan-Manohar mechanism \cite{km}, 
which takes $m_u=0$ at $1/a\sim\Lambda_{\rm\scriptscriptstyle QCD}$, 
and relies on higher-order effects in chiral perturbation theory to simulate $m_u\ne 0$;
this is problematic for several reasons \cite{leut}.

A different class of models posits that $CP$ violation is spontaneous (e.g., \cite{spon}); 
in these models, ${\rm Im}\det m$ is automatically zero at a high scale, but ${\rm Re}\det m$
is not.  Eq.~(\ref{ddetm}) tells us that instanton effects do not contribute to the renormalization of
${\rm Im}\det m$.  Thus, nonperturbative generation of a nonzero up-quark mass 
also enhances the viability of this class of solutions to the strong $CP$ problem. 

We conclude that the effects discussed in \cite{cr} are beneficial, rather than detrimental,
to all versions of the $m_u=0$ solution of the strong $CP$ problem.
 
\begin{acknowledgments}

I thank Michael Creutz for discussions.  
This work was supported in part by NSF Grant No.~PHY00-98395.

\end{acknowledgments}


\begin{thebibliography}{99}

\bibitem{cr} M. Creutz, Phys. Rev. Lett. {\bf 92}, 162003 (2004).

\bibitem{gm} H. Georgi and I. McArthur, Harvard Report HUTP-81/A011 (unpublished).

\bibitem{choi} K. Choi, C. W. Kim, and W. K. Sze, Phys. Rev. Lett. {\bf 61}, 794 (1988).

\bibitem{bns} T. Banks, Y. Nir, and N. Seiberg, hep-ph/9403203.

\bibitem{km} D. B. Kaplan and A. V. Manohar, Phys. Rev. Lett. {\bf 56}, 2004 (1986).

\bibitem{leut} J. M. Gerard, Mod. Phys. Lett. {\bf A 5}, 391 (1990);
H. Leutwyler, Phys. Lett. {\bf B 378}, 313 (1996).

\bibitem{spon} K. Choi, D. B. Kaplan, and A. E. Nelson, Nucl Phys. {\bf B391}, 515 (1993).

\end{thebibliography}
\end{document}